\documentclass[pre,preprint,showpacs,preprintnumbers,amsmath,amssymb]{revtex4}
\usepackage{graphicx}
\usepackage{dcolumn}
\usepackage{bm}
\usepackage[mathscr]{eucal}
\usepackage{mathrsfs}

\newtheorem{Postulate}{Postulate}
\newtheorem{Definition}{Definition}
\newtheorem{Corollary}{Corollary}

\begin{document}

\title{Revisiting nonequilibrium characterization of glass: \\
History dependence in solids}

\author{Koun Shirai$^{1,2}$}
\affiliation{%
$^{1}$Vietnam Japan University, Vietnam National University, Hanoi, \\
Luu Huu Phuoc Road, My Dinh 1 Ward, Nam Tu Liem District, Hanoi, Vietnam
}%
\affiliation{%
$^{2}$SANKEN, Osaka University, 
8-1 Mihogaoka, Ibaraki, Osaka 567-0047, Japan
}%

\begin{abstract}
Glass has long been considered a nonequilibrium material. The primary reason is its history-dependent properties: the obtained properties are not uniquely determined by two state variables alone, namely, temperature and volume, but are affected by the process parameters, such as cooling rates. However, closer observations show that this history dependence is common in solid; in crystal growth, the properties of an obtained crystal are affected by the preparation conditions through defect structures and metallurgical structures. 
The problem with the previous reasoning of history dependence lies in the lack of appropriate specification of state variables. Without knowledge of the latter, describing thermodynamic states is impossible. The guiding principle to find state variables is provided by the first law of thermodynamics. The state variables of solids have been searched by requiring that the internal energy $U$ is a state function. Detailed information about the abovementioned microstructures is needed to describe the state function $U$. This can be accomplished by specifying the time-averaged positions $\{ \bar{\bf R}_{j} \}$ of all atoms comprising the solids. Therefore, $\bar{\bf R}_{j}$ is a state variable for solids. 
Defect states, being metastable states, represent equilibrium states within a finite time (relaxation time). However, eternal equilibrium is nonexistent: the perfect crystal is thermodynamically unstable. Equilibrium states can only be considered at the local level. Glass is thus in equilibrium as long as its structure does not change. 
The relaxation time is controlled by the energy barriers by which a structure is sustained, and this time restriction is intimately related to the definition of state variables. The most important property of state variables is their invariance to time averaging. The time-averaged quantity $\bar{\bf R}_{j}$ meets this invariance property.
\end{abstract}

\maketitle

\section{Introduction}
\label{sec:Introduction}
Glass physics has been studied for over a century but yet lack of consistent understanding persists \cite{Angell95,Rao02,Gotze92,Wolynes12,Berthier11,Stillinger13,Berthier16}. The nature of glass transition, Kauzmann's paradox \cite{Kauzmann48} are such examples. The main problem lies in the difficulty of describing the nonequilibrium characters of glass. Numerous studies on glass have been devoted to the nonequilibrium characters. In this direction of study, the nonequilibrium characterization of glass is taken for granted. 
But it is not obvious why glasses are considered nonequilibrium.
Historically, this nonequilibrium characterization of glass appeared when the third law of thermodynamics was investigated. When the third law was discovered at early of 20th century, it was argued how the residual entropy of glass should be treated, because the residual entropy contradicts the statement of the third law. This contradiction was the subject of debates \cite{Fowler-Guggenheim,Wilson,Lewis-Randall,Wilks,Beattie,Hasse,Landsberg57,Gutzow09}, which continue to this day \cite{Kivelson99,Speedy99,Mauro07,Gupta07,Gupta08, Goldstein08,Reiss09,Gupta09,Moller06,Aji10} \cite{Entropy-20}. The reason for the contradiction remains unsatisfactory, but the generally accepted explanation is that the current state of glass is not in equilibrium but changing toward equilibrium, which was given by Simon \cite{Simon30,Simon51}. Until today, however, no evidence has been found indicating that silica glass is crystallized. On the contrary, considerable evidence suggests that silica glass retains its amorphous state \cite{Berthier16}: the glass articles created in the ancient times have not changed; those created in the dinosaur age are still in the glass state; lunar glass formed more than 3 billion years ago are still in the amorphous state. No metal can retain its original structure for one thousand years. Here we find inconsistency in the usage of equilibrium between glass and usual materials.

Aside from the third law issue, the hysteresis of glass is a common reason for the nonequilibrium characterization of glass.\cite{note-ThirdLawIssue}
The current properties of a glass depend on the past history that the glass underwent. However, closer observations indicate that hysteresis is a common property of solids. It is well known that the mechanical properties of steel are affected by the heat and mechanical treatments in the past. It cannot overstate that the major subjects of metallurgy are concerned with hysteresis behaviors of solids. Regarding hysteresis as an indicator of nonequilibrium will mean that no solids are in equilibrium. 

As shown in the next section, the presence of hysteresis can be restated as the current properties of solids cannot be described by the current values of $T$ and $V$ solely. For gases, it is certain that the state $A$ is uniquely described by these two variables,
\begin{equation}
A(T, V),
\label{eq:StateOfGas}
\end{equation}
when chemical reactions are ignored.\cite{note-svs}
However, the validity of statement (\ref{eq:StateOfGas}) for solids has not been verified. The authors reviewed more than 50 thermodynamics textbooks, but found no answer to this question. The state variables for solids are an open question. This question is alternatively expressed as how the states of solids are specified. In this paper, the word ``state" means the thermodynamics state, which is uniquely specified by state variables regardless of the process by which the current state was obtained.

Bridgman may have been the first to address fully the thermodynamic description of hysteresis in plastic deformation \cite{Bridgman50}. Recognizing that the plastically deformed state cannot be uniquely specified using the state variables of $T$ and stress $\sigma$, he added strain $\varepsilon$ as an independent state variable (in elastic theory, $\varepsilon$ depends on $\sigma$). His idea was valuable for schematic explanations of plastic deformation. However, real deformations are more complicated, and $\varepsilon$ and $\sigma$ in addition to $T$ are insufficient. Kestin stated that dislocation behavior cannot be described well using these variables \cite{Kestin70}. Given the lack of known state variables other than $\varepsilon$ and $\sigma$, a hypothetical quantity called the {\it internal variable}, $\xi(t)$, was introduced as a function of time $t$ to describe hysteresis \cite{Kestin70,Rice71}. An internal variable is dynamic, so hysteresis is considered in the nonequilibrium thermodynamics framework. However, it was soon recognized that a single internal variable cannot adequately describe plastic deformation \cite{Kestin70,Berdichevsky05}. Twenty years after their abovementioned study, Kestin wrote that the problem had not been solved \cite{Kestin92}. More importantly, nobody can say the substance of this internal variable. Kestin quoted Bridgman about internal variables: {\it ``These parameters are measurable, but they are not controllable, which means that they are coupled to no external force variable which might provide the means of control. And not being coupled to a force variable, they cannot take part in mechanical work"} (\cite{Bridgman61}, in Appendix ``Reflections on Thermodynamics"). The current understanding on this point is no more developed than this view made in 1953.

Equilibrium is described using state variables; without determining these variables, one cannot determine whether the state is in equilibrium or not, and speaking of nonequilibrium is illogical. Whether plastically deformed state is in equilibrium or not should be reassessed. The nonequilibrium characterization of glass can be regarded in the same vein. Thus, the issue is not specific to glass; instead, it is a fundamental problem of thermodynamics for all solids. 
Recently, the author has provided a general prescription for this problem \cite{Shirai18-StateVariable}: in addition to $T$, the state variables of a solid are the time-averaged positions, $\{ \bar{\bf R}_{j}\}_{j=1, \dots, N_{\rm at}}$, of atoms that comprise the solid ($N_{\rm at}$ is the total number of atoms). The state $A$ of a solid is fully specified by $\{ \bar{\bf R}_{j}\}$ along with $T$,
\begin{equation}
A(T, \{ \bar{\bf R}_{j} \}).
\label{eq:StateOfSolid}
\end{equation}
In the microscopic theory of solids, this relationship itself is taken for granted, aside from thermodynamics interpretation whether $\bar{\bf R}_{j}$ can be regarded as a state variable or not. In today's material researches, numerous successes have been obtained by density functional theory (DFT), which states that the total energy of a material is uniquely determined by its structure \cite{Parr-Yang89}. 
However, the author's new theory has been criticized by glass researchers due to their deep-seated view that glass is a nonequilibrium material. They claim that numerous published studies on glass support its nonequilibrium characterization. However, no matter how many papers were published, there is no study to trace back to the definition of equilibrium. There is no proof that $T$ and $V$ are only the state variables for solids.
The reasoning of history dependence is not proven for the criterion of equilibrium. From the perspective of the process aspect of hysteresis, hysteresis is indeed a nonequilibrium phenomenon; processes are nonequilibrium by definition. However, in thermodynamics, processes and states are different notions. A nonequilibrium process does not imply that its final state of the process is in nonequilibrium. Unfortunately, this distinction between state and process becomes less clear in the real situations of solids. The main difficulty lies in the involvement of relaxation. Relaxation is an irreversible process, which breaks equality relationships between thermodynamic quantities. Thus, hysteresis should be analyzed in more detail. Accordingly, the thermodynamic characterization of glass should be investigated together with the hysteresis phenomenon, as they are commonly observed in usual crystals. This is one of the purposes of the present study.

The final goal of this paper is to elucidate the equilibrium characterization of glass, which contradicts the generally accepted view. This conclusion is obtained by studying the rigorous definition of state variables for solids, as indicated by Eq.~(\ref{eq:StateOfSolid}). However, for the abovementioned reason, hysteresis of solids should be analyzed before deriving this equation. This is done in Sec.~\ref{sec:Hysteresis}. Hysteresis is observed in a wide range of properties of solids, such as magnetic properties \cite{Chikazumi09,Fiorillo05,Bertotti06}, mechanical properties of metals \cite{Cottrell67,KodaNariyasu73,Suzuki-Takeuchi-Yoshinaga91}, and properties of glass \cite{Nemilov-VitreousState, Hodge94}. The subjects are too broad to be covered in the current study. 
Nonetheless, given the author's materials research background, general investigation about thermodynamics can be provided here. The treated examples are taken from engineering researches. This is because despite the prevalence of history-dependent phenomena in these areas, traditional thermodynamics studies do not highlight them enough. Here, one sees a big estrangement between fundaments of thermodynamics and the engineering applications.
Each area of applications uses different terminologies. The author attempts to keep consistent usage of terminologies and hence they are explained at the beginning, in Sec.~\ref{sec:Preliminary}. This is important when hysteresis is discussed, because the author is aware that there is no rigorous definition of hysteresis in the thermodynamics literature. In fact, hysteresis describes the change in state, but a state cannot be specified without knowing the state variables. Specification of states of solids constitutes the topics of this paper \cite{note-hysteresis-nonEq}.
On analyzing various hysteresis phenomena in solids, it is seen that all metastable states of solids are indeed equilibrium state within the relaxation time. On this basis, a general criterion can be established for equilibrium and specification of states in solids, as shown in Section~\ref{sec:Equilibrium}. The arguments of this section are essentially the same as those in the author's previous preprint \cite{Shirai18-StateVariable}. However, due to the importance of the subject, the arguments are again presented here, with placing particular emphasis on the distinction between static states and very slow processes.

\section{Assumptions and definitions}
\label{sec:Preliminary}
Because we are discussing very fundamental problems, the used assumptions must be declared at the outset. These are the first and second laws of thermodynamics only. Everyone knows these very well. However, in the present arguments, we should be more careful for the expressions than usual. Some explanations are thus provided. Also, the definitions of several terminologies must be given, because they are used differently in different areas.

\subsection{Fundamental Postulates}

\subsubsection{State versus process}
\label{sec:StateProcess}

\begin{Postulate}{{\rm (Internal energy)}}
{The internal energy of a system is a state function,
\begin{equation}
U = U(T, \{ X_{j} \}),
\label{eq:stateU}
\end{equation}
where $X_{j}$ denotes state variables of the system other than $T$.}
\label{PS:internal-energy}
\end{Postulate}

\noindent
This means that $U$ is determined by the current state only but does not depend on the process in which the current state was obtained. 
This expression is equivalent to the usual expression of the first law of thermodynamics (\cite{Fermi}, p.~21), which states the energy conservation between $U$, mechanical work $W$, and heat $Q$ as
\begin{equation}
dU = W + Q.
\label{eq:first-law}
\end{equation}
We can regard Eq.~(\ref{eq:first-law}) either as the definition of heat $Q$ given Postulate I (as adapted by Fermi  \cite{Fermi}) or as the definition of $U$ if $Q$ is already defined.
There is no proof for this Postulate, because it is a fundamental law of physics (\cite{Gyftopoulos}, p.~31); a fundamental law cannot be deduced from other laws. Only contradiction with an experiment will disprove this law.
By changing variable $T$ with entropy $S$, Eq.~(\ref{eq:stateU}) turns to be the fundamental equation \cite{Callen},
\begin{equation}
U = U(S, \{ X_{j} \}).
\label{eq:FRE}
\end{equation}
This name was used by Gibbs more than a century ago, predating any atomic theory \cite{Gibbs-TD}. Today, the name of {\em the fundamental relationship of equilibrium} (FRE) is better by considering the meaning \cite{Gyftopoulos}. Both Eqs.~(\ref{eq:stateU}) and (\ref{eq:FRE}) are the FRE. 
The FRE contains anything about equilibrium properties of a system. The problem is which variables $X_{j}$ are needed to specify $U$. 

\begin{figure}[htbp]
  \centering
    \includegraphics[width=120 mm, bb=0 0 1024 768]{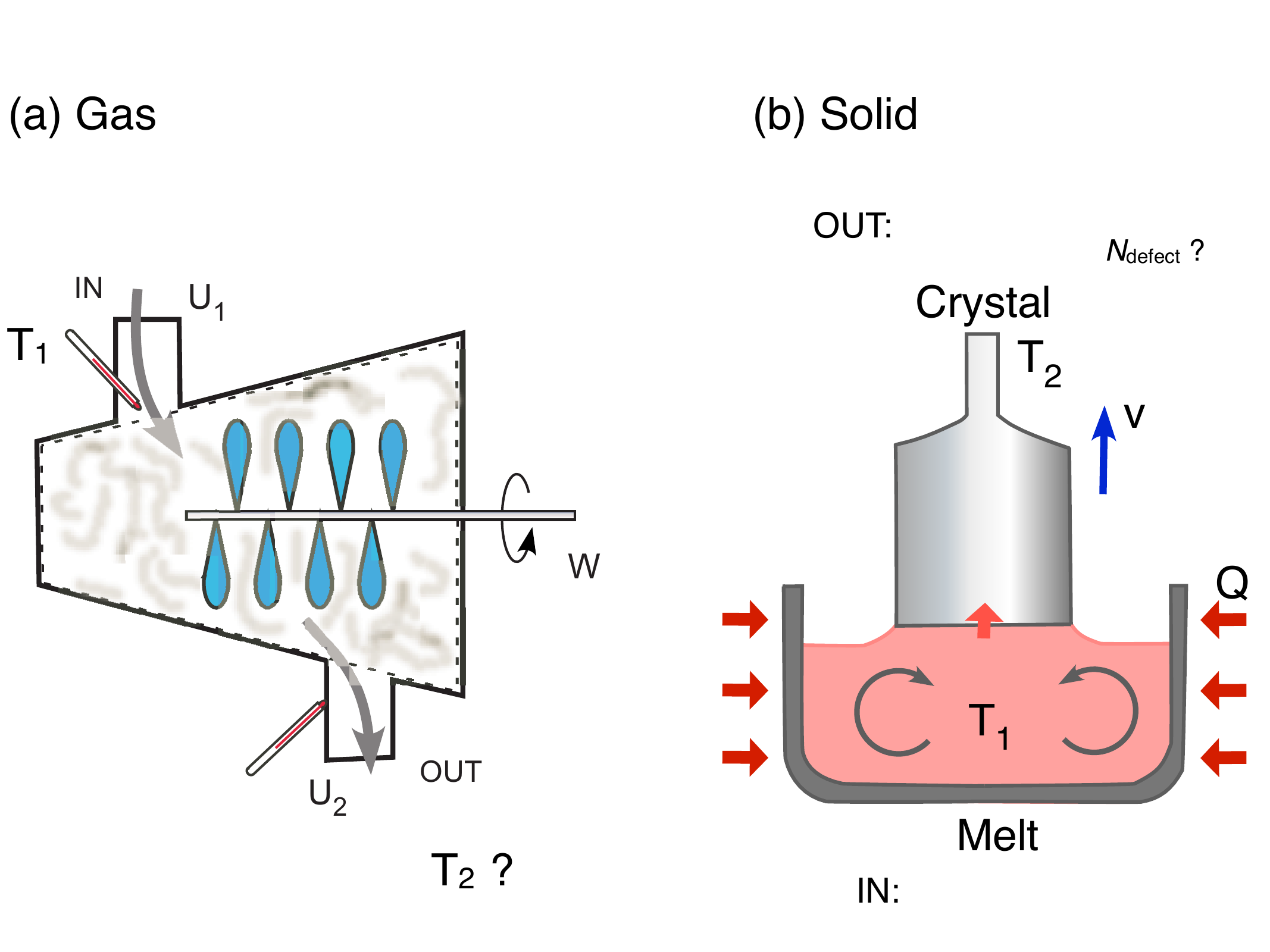} 
  \caption{(a) Gas turbine. Hot vapor flows in the turbine at $T_{1}$ and flows out at $T_{2}$, while generating work $W$. (b) Crystal growth from a melt at $T_{1}$ at the pulling rate $v$.
  } \label{fig:Turbine-CrystalGrowth}
\end{figure}

Although the author is not aware of published articles proclaiming that $U$ of glass does not accord Postulate I, he often heard someone's claiming in a speaking manner that $U$ is not well defined for glasses, because the same glasses with different properties are obtained by changing the the cooling rate. At this point, we see that missing variables leads to a serious mistake. The notions of state and process must be distinguished in thermodynamics.
Let us consider a turbine generating electric power, as shown in Fig.~\ref{fig:Turbine-CrystalGrowth} (a). Hot vapor flows into the turbine, with $T_{1}$ being the inlet temperature. Wasted water flows out of the turbine, with $T_{2}$ being the outlet temperature. Let us ask what is the value $T_{2}$ for a given value $T_{1}$. There is no unique answer. It depends on the process throughout which the water is underwent. Inside the turbine, complicated processes of energy exchanges occur between the vapor and the blades. Some heat is lost due to friction between water and any part of the turbine. The resulting properties of the rejected water depend on the process. Nonetheless, the state of the water can be uniquely determined using the outlet $T$ and $V$ values after thermodynamic equilibrium is recovered there. The details of the process are unnecessary. Everybody understands this distinction for gases. However, when we investigate solids, everything becomes clouded, because we are not sure which are the state variables that uniquely specify the current state of a solid.

In the glass case, the fact that glasses with different properties are obtained by different histories merely means that different {\em states} of the glass are obtained depending on the preparation conditions. The critical issue is whether the current value of $U$ can be specified by adding extra variables $Z$ in addition to $T$ and $V$,
\begin{equation}
U = U(T, V, Z).
\label{eq:U=TVZ}
\end{equation}
If we find $Z$ being a state variable, it is consistent with Postulate I. We conclude that any state $A$ of the system can be specified by a set of state variables $(T, V, Z)$ regardless of the past history.
If we find $Z$ being not a state variable, similar to heat and work, the only conclusion is that it violats Postulate I. 
Bridgman does not recommend to rush to conclude that the laws of thermodynamics do not apply to our system, if at first we are unsuccessful in finding a set of parameters which determine an energy function (\cite{Bridgman61}, p.~59). Instead, he suggests to suspect that we have not a complete list of parameters of state.

A good suggestion is gained from the study on defects in crystals. Figure \ref{fig:Turbine-CrystalGrowth} (b) shows a crystal growth from a melt. The defect concentration depends on the pulling rate of crystal (the subject is discussed more details in Sec.~\ref{sec:DefectConcentration}). Thus, the properties of the obtained crystal have history dependence in the above sense. Nonetheless, researchers studying defect properties do not need to know the process by which these defects were obtained. After identifying the defect structure, they can relate the properties of the defect to its structure. The defect energy depends on the details of atom positions, although these are time-averaged positions, as
\begin{equation}
U = U(T, \{ \bar{\bf R}_{j} \}).
\label{eq:UofSolids}
\end{equation}
Defects with the same structures are expected to have the same behavior despite being obtained through different processes.
This equation can be the FRE for solids. This functional relationship is taken for granted in microscopic studies in materials research. In DFT, through Born--Oppenheimer approximation, the total energy $E(t)$ of a solid at a time $t$ is obtained using the sum of the kinetic and potential energies as
\begin{equation}
E(t) = \frac{1}{2} \sum_{j} M_{j} v_{j}(t)^{2} + E_{gs}( \{ {\bf R}_{j}(t) \} ),
\label{eq:BOapprox}
\end{equation}
where $M_{j}$ and $v_{j}$ are the mass and velocity, respectively, of $j$th atom and $E_{gs}$ is the ground-state energy of electrons for the current atom positions. Equation (\ref{eq:BOapprox}) holds not only when system is in equilibrium, but also when the system is at any state such as a saddle point in the adiabatic potential. The internal energy $U$ in thermodynamics is the time-averaged microscopic total energy $E(t)$ of a system in equilibrium, $U=\overline{E(t)} $. When the system is in equilibrium, the instantaneous atom position is expressed as the sum of the equilibrium position $\bar{\bf R}_{j}$ and its small displacement ${\bf u}_{j}(t)$ from the equilibrium position as
\begin{equation}
{\bf R}_{j}(t) = \bar{\bf R}_{j} + {\bf u}_{j}(t).
\label{eq:SmallDisplacement}
\end{equation}
By expanding $E_{gs}( \{ {\bf R}_{j}(t) \} )$ with respect to the small displacement ${\bf u}_{j}$ around the equilibrium position $\bar{\bf R}_{j}$, we obtain
\begin{equation}
U = \overline{E(t)} = E_{\rm st}(\{ \bar{\bf R}_{j} \}) + \frac{1}{2} \sum_{k} \omega_{k}^{2} \overline{q_{k}^{2}},
\label{eq:UofSolids1}
\end{equation}
where $q_{k}$ and $\omega_{k}$ are $k$th normal coordinate and its frequency, respectively \cite{Shirai20-GlassState}. $E_{\rm st}$ is the structural part of the total energy. Equation (\ref{eq:UofSolids1}) turns out to be equivalent to the general expression (\ref{eq:UofSolids}). The $T$ dependence in Eq.~(\ref{eq:UofSolids}) stems from the $T$ dependence of $q_{k}$ through the Bose factor on the harmonic approximation. Beyond this approximation, the structural part $E_{\rm st}(\{ \bar{\bf R}_{j} \})$ comes to participate the $T$ dependence \cite{Shirai22-SH}.
Comparison of Eq.~(\ref{eq:UofSolids}) and the formal expression (\ref{eq:stateU}) shows that the equilibrium positions of atoms $\{ \bar{\bf R}_{j} \}$ can be considered state variables, as is found in \cite{Shirai18-StateVariable}. 
This heuristic finding that $\{ \bar{\bf R}_{j} \}$ are state variables for solids is reformulated in a general manner in Sec.~\ref{sec:Equilibrium}. However, the significance of the relationship (\ref{eq:UofSolids}) for describing the properties of solids should be recognized through real examples, particularly because the history-dependence character of glass is a common property across all solids. This is the purpose of Sec.~\ref{sec:Hysteresis}.

\subsubsection{Equilibrium condition}
\label{sec:SecondLaw}
Regarding definition of equilibrium, the most important guiding principle is the second law of thermodynamics. There are many expressions for the second law of thermodynamics: namely, 21 different ways according to \cite{Capek}. The most recent one, given by Gyftopoulos and Berreta, is

\begin{Postulate}{{\rm (The second law of thermodynamics)}}
There is one and only one equilibrium state for a set of constraints $\{ \xi_{j} \}$ and a given $U$.
{\rm (Ref.~\cite{Gyftopoulos}, p.~63) } 
\label{PS:second-law}
\end{Postulate} 

\noindent
Someone may have, at the first glance, strange impression in this expression: why is existence of equilibrium needed for stating the second law? However, it turns out that this is equivalent to the maximum entropy principle, which is generally accepted: a system attains the maximum of entropy at equilibrium (\cite{Callen}, p.~103). The maximum entropy principle was proposed more than a century ago by Gibbs \cite{Gibbs-TD}. He considered the general criteria for equilibrium and stated:

\vspace{3mm} \noindent
{\bf Gibbs' criterion for equilibrium}:
{\em For the equilibrium of any isolated system it is necessary and sufficient that in all possible variations of the state of the system which do not alter its energy, the variation of its entropy shall either vanish or be negative.} 
{\rm (Ref.~\cite{Gibbs-TD}, p.~56) }

\vspace{2mm}
\noindent
Actually, this statement is equivalent to Postulate \ref{PS:second-law}. Hence, this statement can be used as either the statement of the second law or as the definition of equilibrium, provided entropy is already defined. However, readers may be aware of attachment of constraints in Postulate \ref{PS:second-law}, while no such conditioning is seen in the Gibbs' statement. This point is crucial to this study and is fully discussed in Sec.~\ref{sec:Equilibrium}.

\subsection{Terminologies}
\label{sec:Terminology}

\paragraph{Structure.} 
\label{sec:Structure}
Any crystal has multiple structures. Silicon crystal normally has a diamond structure, but it has a hexagonal form at high pressures. In the crystallographic literature, a crystal structure is specified by the type of its unit cell (Bravais lattice) plus the basis atoms in a unit cell. In this paper, we use the word ``structure" in more general way to mean any atom arrangement, including defect states. Atom arrangement is specified by the time-averaged positions $\{ \bar{\bf R}_{j} \}$ \cite{note-timeaverage}.

Displacing one atom to an interstitial site creates a different structure from the original one. Many defect structures exist, all of which are specified by the atom position, $\bar{\bf R}_{j}$. The atoms in the dislocation are displaced by the Burgers vector. Similarly, stacking fault, vacancy, and voids are expressed by the displacements of atom positions. Domain, grain, and twin structures, and even macroscopic inhomogeneous deformations can also be described by time-averaged atom positions, $\bar{\bf R}_{j}$. 
When need arises to distinguish from the crystallographic structure, the collective words microstructure or configuration are used to refer to different atom arrangements $\{ \bar{\bf R}_{j} \}$.
All these structures can be described using the set $\{ \bar{\bf R}_{j} \}$. As we already know that $\{ \bar{\bf R}_{j} \}$ presents a set of state variables, structures of a solid are specified by state variables. Therefore, all existing structures are equilibrium states, which are specified by a set of state variables $(T, \{ \bar{\bf R}_{j} \})$, as in Eq.~(\ref{eq:StateOfSolid}).
Without information of $\{ \bar{\bf R}_{j} \}$, the states of a solid cannot be specified. Readers will be confirmed that the specification (\ref{eq:StateOfSolid}) is reasonable, as reading following arguments.

The structure is defined only when each atom has a definite average position $\bar{\bf R}_{j}$. A set of instantaneous positions $\{ {\bf R}_{j}(t) \}$ in Eq.~(\ref{eq:SmallDisplacement}) does not constitute a structure. According to this definition, liquids have no structure, because their constituent atoms have no unique average position. The distinction between the time-averaged and instantaneous position of atoms is important to this extent (in Sec.~\ref{Sec:StateVariable-Def}).

\begin{figure}[ht!]
\centering
\includegraphics[width=.45 \textwidth, bb=0 0 450 280]{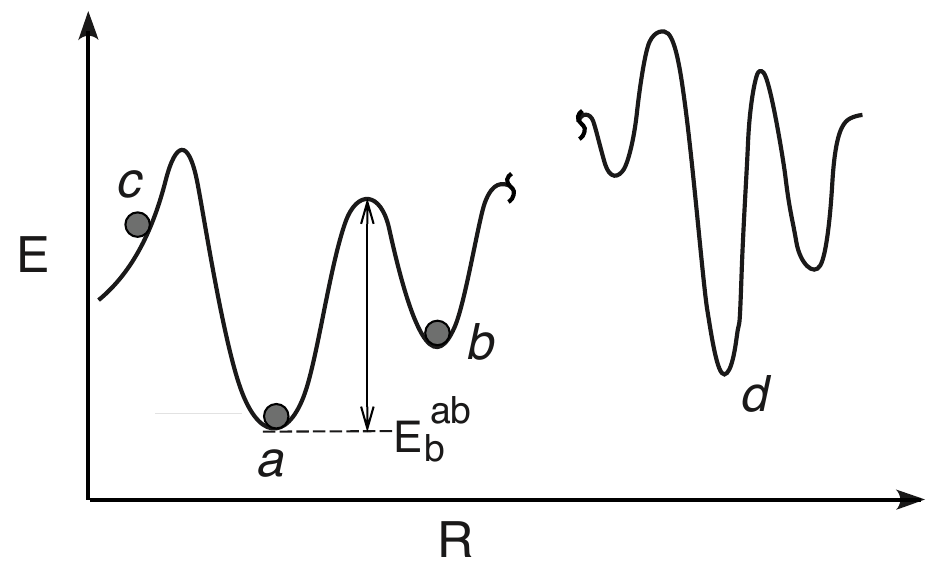}
\caption{
Energy landscape of a solid. Structure $a$ is the lowest-energy state, whereas $b$ and $d$ are metastable states. In the glass literature, a frozen state is identified by $c$. The atom positions $\{ \bar{\bf R}_{j} \}$ are collectively expressed by ${\bf R}$. The time-averaged position $\bar{\bf R}$ is defined only for the points of local minima. State $d$ is located far distant from $b$ in the configurational space ${\bf R}$. The energy barrier built around state $d$ is the largest, although the energy of $d$ is not the lowest.
}
\label{fig:meta-frozen} 
\end{figure}

\paragraph{Stable, metastable, and frozen states.} The distinction between stable state (global equilibrium) and metastable state (local equilibrium) is often useful in materials research. In Fig.~\ref{fig:meta-frozen}, state $a$ presents the lowest-energy state and is accordingly the global minimum state, whereas state $b$ is a metastable state. However, in the thermodynamics context, their distinction is meaningful only in a relative sense. In the normal conditions, the stable phase of hydrogen is the gas state of hydrogen molecule H$_{2}$: the lowest-energy state. Considering nuclear reactions, a helium nucleus is by far lower in energy than a H$_{2}$ molecule. By observing successive nuclear reactions, we find that an iron nucleus is the most stable one. Hence, it is difficult to determine the lowest-energy state without restricting the range of searching. Furthermore, in thermodynamics, the stability condition is given by the free energy $F$ rather than energy. The minimum-$F$ state changes with $T$ and other conditions. In Sec.~\ref{sec:DefectConcentration}, even if nuclear reactions are disregarded, we will see that equilibrium can only be considered at the local level.

In the glass literature, special word {\it frozen state} or {\it kinetically frozen state} is used to refer to the glass state. The meaning is explained that, for glass-forming materials, the viscosity of the liquid state is so high that atom motions cease before the equilibrium state is reached. Someone regards the frozen state as a metastable state like state $b$ in Fig.~\ref{fig:meta-frozen}. In this case, frozen states do not differ from metastable state. Hence, some others explain that the frozen state is even not metastable state but an unstable state such as $c$ in the figure (this explanation is found in (\cite{Wilks}, p.~61). Similar to a stone on a slope, the structure $c$ is fixed due to the frictional force. However, the idea of frictional force does not exist in quantum mechanics. Friction is meant the energy dissipation process in which the energy initially assigned to a particular freedom is released to an uncontrollably large number freedoms. Energy barriers is the substance that impede atom motion. In this paper, frozen states are regarded as metastable states. Also, the megabasin or intrinsic structures used in the glass literature are avoided. In the present context, all states are local minima in the potential landscape in the configurational space.

\paragraph{Transition between structures.} 
\label{sec:TransitionRate}
The stability of a structure is often determined not by its depth in energy but by the energy barrier $E_{b}$, which prevents the current structure from transitioning to other ones. Nuclear fission is caused by irradiation of thermal neutrons (the energy at room temperature), whereas nuclear fusion occurs only at very high temperatures (at least $10^{8}$ K). A theory on the transition rate $\Gamma$ has been established based on quantum mechanics \cite{Eyring64}. The rate $\Gamma_{ab}$ of transition $a \rightarrow b$ is given as
\begin{equation}
\Gamma_{ab} = \nu_{0} \exp \left( -\frac{E_{b}^{ab}}{k_{\rm B}T} \right),
\label{eq:rate}
\end{equation}
where $E_{b}^{ab}$ is the energy barrier between states $a$ and $b$ and $\nu_{0}$ is the attempt frequency. The inverse of this transition rate is called the mean transition time, or the {\em relaxation time}.
The transition rate, Eq.~(\ref{eq:rate}), focuses on a specific type of transition $a \rightarrow b$. Normally, a given state $a$ has many terminal states $b'$ into which $a$ can transform. When we are interested in how long the initial state $a$ is retained, we use the {\em lifetime} of state $a$, which is determined by the shortest time among many transitions $\{a \rightarrow b'\}$. In most cases, the lifetime and relaxation time are used interchangeably.

Transitions described by Eq.~(\ref{eq:rate}) occur in a wide range of phenomena, such as chemical reactions, solid-state diffusion, and electrical conductivity, in the form of the Arrhenius law or thermal activation. The most important feature of the thermal activation is its exponential dependence on time $t$ and on the inverse of $T$. Sometimes, however, we observe deviations from this exponential dependence in real experiments \cite{Williams55,Moynihan74,Hodge94}.
The rate equation (\ref{eq:rate}) refers to an elemental process with a single barrier $E_{b}$. Hence, for a transition consisting of a series of energy barriers, the observed quantity may not obey the Arrhenius law. 

\begin{figure}[htbp]
    \centering
    \includegraphics[width=75 mm, bb=0 0 450 280]{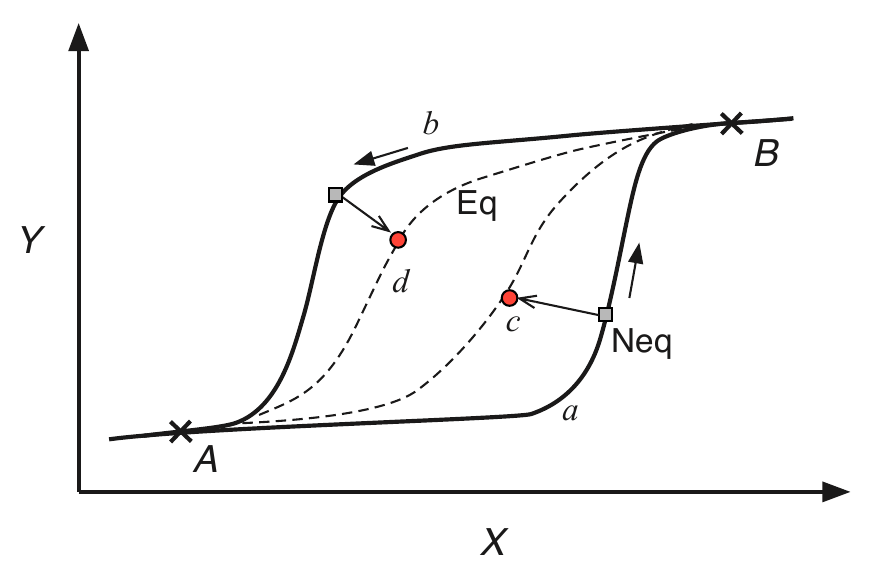} 
  \caption{Hysteresis loop. Input $X$ and its response $Y$ of a system return to the initial values upon a cycle with different paths for the forward and backward directions of $X$.
  } \label{fig:Hysteresis}
\end{figure}

\paragraph{Hysteresis.} 
\label{sec:DefHysteresis}
Solids exhibit many history-dependent phenomena. Specific names, such as aging, rejuvenation, and memory effect, are used depending on their application. In this paper, these phenomena are all referred to as hysteresis. 
Thermodynamics textbooks rarely give a rigorous definition of hysteresis \cite{Brokete96,Morris11,Muschik93}. Probably, this is because hysteresis is generally considered a nonequilibrium phenomenon. Hysteresis is best known as a hysteresis loop in magnetic materials \cite{Chikazumi09,Fiorillo05,Bertotti06}. It also occurs at plastic deformation \cite{Bridgman50,Kestin70} and at the glass transition \cite{Nemilov-VitreousState}. In Sec.~\ref{sec:Hysteresis}, we will see that hysteresis is a common response of solids.
A hysteresis loop is shown in Fig.~\ref{fig:Hysteresis}. For a cyclic change of an external field $X$, the response of a material property $Y$ exhibits a loop. 
Mathematically, hysteresis is expressed as a multivalued function $Y=Y(X)$ \cite{Brokete96}. However, this understanding is unsatisfactory in the thermodynamics context. In a Carnot cycle, known as the best heat cycle, setting the input to $X=V$ and the output to $Y=P$ leads to a hysteresis loop. In thermodynamics, all thermodynamic functions are expressed as multivariables (not multivalued) functions, so hysteresis in this sense always occurs. Attaching the name hysteresis in this manner is useless.

Let us investigate in more detail the meaning of history dependence on a hysteresis loop.
In Fig.~\ref{fig:Hysteresis}, the property $Y$ returns to its initial value after a cyclic change in the input $X$. However, this does not guarantee the perfect restoration of the initial state $A_{i}$. Experimentalists identify the initial state $A_{i}$ by measuring the input $X_{i}$ and the response $Y_{i}$. However, other state variables $Z$ may exist. The state is fully specified by them as $A_{i} = A(X_{i}, Y_{i}, Z_{i})$. Then, the final state $A_{f}$ could be $A_{f} = A(X_{i}, Y_{i}, Z_{f})$, where $Z_{f} \neq Z_{i}$. 
This is the case of the specific heat ($C_{p}$) characteristic of glass. For glass transition, a $C_{p}-T$ curve across the glass-transition temperature, $T_{g}$, exhibits a hysteresis loop. When a heat cycle is started from the glass state, even though the output $Y=C_{p}$ returns to the initial value, another property enthalpy $H$ does not generally restore the initial value: a circle integration $\oint C_{p} dT$ does not vanish (\cite{Shirai20-GlassState}, p.~9).
In the steady-state operation of a magnetic motor, the final state $A_{f}$ of the iron core seems to restore the initial state $A_{i}$, if a long-term degradation is ignored \cite{Bertotti06}. However, in this case, the input electromagnetic work is dissipated for heat generation \cite{Maugin91}. It gives an impact on the environment. Therefore, we can define hysteresis preferably as follows.
\begin{Definition}{\rm (Hysteresis)}
Hysteresis is a process where the initial state $A_{i}$ of the system is not recovered after one cycle of the external field $X$ without causing any change in the environment. 
\label{def:hysteresis}
\end{Definition} 
This shows that hysteresis is essentially synonymous with irreversible processes. The former is a special case of the latter, when an external field $X$ is changed in a cyclic manner. The main problem for glass is how we identify restoration of the initial state. Even we are not sure what the same state of a glass means.

\paragraph{Time dependence versus hysteresis.} 
\label{sec:RateDep}
The approach of thermodynamics is to study the properties of a system by its static states. 
History dependence implies time dependence. By definition, a process describes time dependence and the state during a process is a nonequilibrium state. 
However, history dependence is not necessarily the same as time dependence in the thermodynamics context. This is a confusing point, as already described in Sec.~{\ref{sec:StateProcess}}. 
When we speak of the history dependence, in most cases, we mean the effects of previous treatment on the current properties after the treatment process was finished. Usually, the measurement of current properties is performed at sufficiently long time after the treatment, with ensuing that the current state is static. Static state is an equilibrium state; this equivalence is discussed in Sec.~\ref{Sec:PassiveResistance}. Normal experiments of annealing and mechanical treatments belong to this category. 
On the contrary, hysteresis loops are normally obtained during time development of an external field $X(t)$, as in Fig.~\ref{fig:Hysteresis}. In this case, both the external filed $X(t)$ and the material response $Y(t)$ simultaneously change. Relaxation processes are superimposed in the $X(t)-Y(t)$ curve, which complicates thermodynamic analysis. This complication, however, can be eliminated by stopping the sweeping of external field at a fixed value $X_{0}$. The system property $Y(t)$ continues to change for a while even after stopping the change in $X$. This change is purely due to a relaxation effect. But this relaxation change eventually finishes, bringing about a static state for the system. This static state is an equilibrium state, and is one of metastable states that the system can take. This relaxation time, $\tau_{Y}$, depends on the property $Y$ that we observe. 
The above process of stopping and waiting is repeated at every value $X$, and finally the equilibrium (static) curves ($c$ and $d$ in Fig.~\ref{fig:Hysteresis}) are obtained. In this paper, we eliminate time dependence in this manner to investigate hysteresis.

\paragraph{Separation of thermal and mechanical relaxations.} 
\label{sec:ThermalRelax}
The above argument about static curves highlights the need to determine the waiting time needed to reach the static state. A solid has many relaxation times $\tau_{j}$ associated with $j$th freedom of motions, Eq.~(\ref{eq:SmallDisplacement}).
The most important condition for equilibrium is the presence of temperature $T$. This is clear for gas systems. In turbulence of a gas, $T$ is not defined; if it is defined locally but not uniformly distributed, then the system will be in nonequilibrium. By contrast, for solids, the temperature of the solid is usually easy to measure even for history-dependent phenomena. Let us call this situation---the temperature is well defined---{\it thermal equilibrium} \cite{Shirai20-GlassState}. For the total equilibrium (thermodynamic equilibrium), constancy of other state variables $X_{j}$ is required. This part is referred to as {\it mechanical equilibrium}. Again, need arises to specify which are state variables $X_{j}$: if we do not know which are state variables, it is illogical to say nonequilibrium.
We have already seen that a reasonable choice of the state variables for solids is a set of $\{ \bar{\bf R}_{j} \}$, which determines structure $K$. Hence, the mechanical relaxation means the relaxation of $K$th structure, whose relaxation time is $\tau_{K}$.

In a previous study on glass transition \cite{Shirai20-GlassState}, the author clarified the decoupling between the thermal and mechanical relaxations, because the glass transition is such a slow process.
The atom position in a solid is broken up as the right-hand side of Eq.~(\ref{eq:SmallDisplacement}). The second term ${\bf u}_{j}(t)$ on this side represents phonons, which carry the fast motions. Its relaxation time $\tau_{T}$, which is called thermal relaxation time, is of the order of 10 ps. This time is so short that a waiting time in usual experiments is not needed to end the thermal relaxation. The time average of the phonon contribution to $U$ is determined by $T$, and thus is not an independent state variable. 
The first term $\{ \bar{\bf R}_{j} \}$ carries the mechanical relaxation. In the glass transition, $\tau_{K}$ is on a macroscopic timescale of an order of 1 s. This large separation between $\tau_{T}$ and $\tau_{K}$, namely,
\begin{equation}
\tau_{K} \gg \tau_{T},
\label{eq:AdiabaticSeparation}
\end{equation}
is called the {\em adiabatic separation of the second kind} in the previous study \cite{Shirai20-GlassState}. In this case, even during the transition process, an instantaneous state $A(t)$ can be specified using a set of instantaneous state variables $( T(t), \{ \bar{\bf R}_{j}(t) \})$. The time average of $\bar{\bf R}_{j}(t)$ is taken over a period $t_{0}$ much longer than $\tau_{T}$ but much shorter than $\tau_{K}$ in order to resolve the time development $\bar{\bf R}_{j}$,
\begin{equation}
\tau_{K} \gg t_{0} \gg \tau_{T}.
\label{eq:ASofGlass}
\end{equation}
On this timescale, the hysteresis in the $C_{p}-T$ curve can be described on thermodynamics methods, with help of the relaxation time \cite{Shirai21-GlassHysteresis}.
Crystalline tin has two allotropes: $\alpha$-Sn is the low-temperature phase, and $\beta$-Sn is the high-temperature phase, with the transition temperature being $T_{\rm tr} = 13^{\circ}$C. Experimentalists, by some means, have successfully retained $\beta$-Sn at low temperatures, even $T \approx 0$K: the vanishing entropy difference of these two phases was proven in this way (\cite{Wilson}, p.~202).
When $\beta$-Sn survives at slightly lower than $T_{\rm tr}$ in a cooling process, the measurement on $\beta$-Sn should be quickly finished within the structural relaxation time $\tau_{\rm \beta-Sn}$ in order to meet the condition (\ref{eq:ASofGlass}). On the contrary, the measurement on $\alpha$-Sn is easy. The waiting time $t_{w}$ is not long, and the condition $t_{w} \gg \tau_{\rm \beta-Sn}$ is easily met.
However, because of the exponential form of the $T$ dependence of the relaxation time (Eq.~(\ref{eq:rate})), $\tau_{\rm \beta-Sn} \rightarrow \infty $ as $T$ approaches 0 K. Measurement on $\beta$-Sn is easy, whereas measurement on $\alpha$-Sn in this procedure becomes almost impossible.

The waiting time $t_{w}$ needed to obtain static states for a hysteresis loop depends on the property $Y$ by which the hysteresis loop is observed. With the relaxation time of $Y$ denoted as $\tau_{Y}$, this condition is expressed as
\begin{equation}
t_{w} \gg \tau_{Y} \gg \tau_{T}.
\label{eq:conditionQS}
\end{equation}
This condition is not a serious restriction for magnetic hysteresis. Magnetic motors exhibit virtually no rate dependence in normal operation because of the fast magnetic response \cite{Bertotti06}.
If the condition (\ref{eq:conditionQS}) is met, a hysteresis curve $Y-X$ presents a static property, $Y=Y(X)$, although both variables are measured during time variation. In this study, all hysteresis phenomena are regarded to be within this static limit. Accordingly, it should be noted that this work is not an extended thermodynamics study of irreversible processes (\cite{deGroot-Mazur,Prigogine67} and also note \cite{note-hysteresis-nonEq}), which describe time evolution of nonequilibrium states. This work remains in the classical framework; only static states are considered.

\section{History dependence in solids}
\label{sec:Hysteresis}

\subsection{Phase diagram}
\label{sec:PhaseDiagram}
It is appropriate to begin our argument from discussing phase diagram, because phase diagrams is the most suitable tool to investigate equilibrium states of materials. For single-component systems, a phase diagram provides a map of regions of equilibrium phases in the space of $T$ and $P$. For multi-components systems, it projects equilibrium regions on the space of $T$ and the chemical compositions $x$. Phase diagrams are useful for material researchers to find suitable conditions of preparation. However, in real preparation, the obtained phase often differs from that predicted by the phase diagram. For a steel (Fe-C) system, the $\gamma$ (austenite) phase is the stable phase at high temperatures, where C atoms are uniformly dispersed. When a C-containing $\gamma$ phase is cooled, different phases are obtained, depending on the cooling rate \cite{Cottrell67}. The hardest phase of martensite ($\alpha$-Fe), where C atoms are located at interstitial sites, is obtained by high cooling rates. Pearlite phases, which are multi-phases of $\alpha$-Fe and cementite (Fe$_{3}$C), are obtained by slow cooling rates. Although the obtained pearlite structures are crystallographically identical, they have different layered textures, resulting in different hardness. 

The reason for this rate dependence is the time delay for atom rearrangement. The interstitial C atoms in martensite must move to an appropriate site in cementite, which is stable at low temperatures. However, an energy barrier lies between these two sites. Two configurations of C-dissolved $\alpha$-Fe and pearlite are represented by local minima ($a$ and $b$ in Fig.~\ref{fig:meta-frozen}). Atom migration in solids is so slow that interstitial C atoms cannot have enough time to reach an appropriate site of Fe$_{3}$C at fast cooling rates. This time delay in atom rearrangement is common for solids, although the magnitude of delay varies between material. This time delay causes the path dependence in the final product: various pearlite structures form through metastable phases of C-dissolved $\alpha$-Fe. 

Although the possibility of having metastable intermediate phases is by far high for alloy systems because of their large number of possible configurations, this also happens in single-component systems, even without impurities.
It is generally accepted that the phase diagram of boron has four allotropes \cite{Shirai17-review}: $\alpha$-rhombohedral, $\beta$-rhombohedral, $\gamma$-orthorhombic, and $\alpha$-tetragonal phases. Until now, however, more than 10 phases were reported. Although some may be misidentified due to, for example, twin structures, experimentalists commonly acknowledge that various intermediate phases appear during synthesis; these intermediate phases do not appear in the phase diagram \cite{Cottrell67}. The dependence of transition paths through intermediate states is formulated as the Ostwald's step rule. 
Despite being the lowest-energy $\delta$-orthogonal phase among $\alpha$-tetragonal derivatives, it did not appear in the phase diagram, where only $\alpha$-tetragonal phase is present at high temperatures \cite{Uemura19}. The reason is that, although $\delta$-orthogonal phase is the lowest-energy state at low temperatures, $\alpha$-tetragonal is stable at high temperatures, which blocks creation of $\delta$-orthogonal phase by high-temperature synthesis. Nonetheless, an effective transition path to $\delta$-orthogonal phase was found by help of hydrogenation/dehydrogenation \cite{Uemura19}.

Probably, diamond is the most famous example that does not obey phase diagram. Diamond is by the most stable material on the earth. Yet, this phase does not occupy any stable $T$ region in the phase diagram of carbon unless extreme high pressure is applied: graphite is stable over all the $T$ range at normal pressures \cite{Bundy96}. This example evidences that stability of materials is rather controlled by the energy barrier, as explained in II B$c$. In Fig.~\ref{fig:meta-frozen}, a metastable state $d$ has a higher energy than $a$. But the energy barrier surrounding $b$ is larger than that around $a$.
The peculiarity of diamond extends beyond this point. Until the 1980s, researchers believed that the high-pressure method was only the method for synthesizing human-made diamond. Surprisingly, diamond can now be synthesized even at low pressures through plasma chemical-vapor deposition (CVD) method \cite{Spitsyn81,Kamo83,Kawarada97}. The quality of the thin-film diamond synthesized by CVD is surprisingly good to be able to achieve heavily doping \cite{Takano04}. Because diamond is not the lowest-energy state at normal pressures, the mechanism must be beyond the Ostwald step rule. The involvement of hydrogen gas in plasma processes may play an active role in blocking the transition path to graphite: see \cite{Wakatuki88} for review.
Phase diagram indicates how the equilibrium state is changed by changing $T$ and $V$, but it does not say that $T$ and $V$ are the only state variables for solids. 

\subsection{Equilibrium concentration of defects}
\label{sec:DefectConcentration}

\begin{figure}[htbp]
    \centering
    \includegraphics[width=75 mm, bb=0 0 650 420]{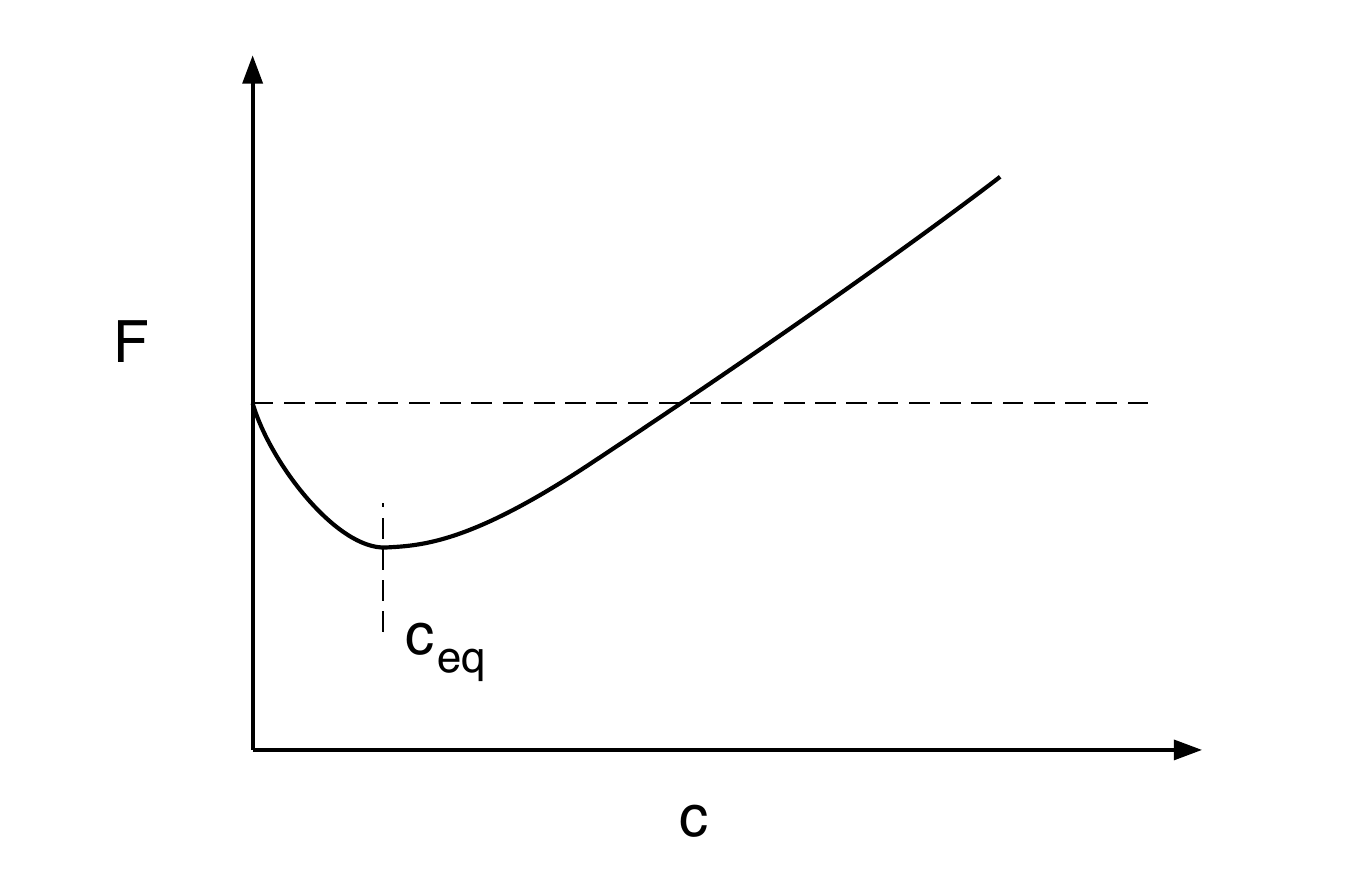} 
  \caption{Free energy $F$ of defect state as a function of the defect concentration $c$.
  } \label{fig:F-c}
\end{figure}

We have seen that a phase diagram presents only fractions of existing phases. More importantly for the present study, the same phase exhibits different properties depending on microstructures. 
All crystals have a certain amounts of defects. Even without foreign atoms, crystals have intrinsic defects, such as vacancies and interstitials. We initially limit our argument to intrinsic defects for simplicity. The properties of an obtained crystal are affected by defects. Theoretically, defects have an equilibrium concentration, $c_{d}$, which is a function of $T$ and expressed as
\begin{equation}
c_{d}(T) = c_{0} \exp \left( -\frac{E_{d}}{k_{\rm B}T} \right),
\label{eq:cofdefect}
\end{equation}
where $E_{d}$ is the formation energy of a given species of defects and $c_{0}$ is the concentration of the available sites for this species.
Some researchers claim that defect states are not equilibrium states and accordingly should not be considered in thermodynamics. Although the nonlowest energy state is true because of positive $E_{d}$, equation~(\ref{eq:cofdefect}) is a consequence of thermodynamics investigation, as the term ``equilibrium concentration" suggests. It is obtained by minimizing the free energy $F(c)$ with respect to the defect concentration $c$. $F(c)$ has an entropic term $-k_{\rm B}T c \ln c$. At $c=0$, $F(c)$ has a negative slope with respect to $c$, as shown in Fig.~\ref{fig:F-c}. This says that the perfect crystal is not even a local minimum state in the thermodynamics context; in fact, no perfect crystal exists.

In practice, one never observes achievement of the exact equilibrium concentration given by Eq.~(\ref{eq:cofdefect}) in real experiments. The defect concentration is rather governed by the preparation conditions. At the melting temperature $T_{m}$, where a sample is prepared, vacancies are introduced. The concentration $c_{d}$ could be the equilibrium concentration $c_{d}(T_{m})$ at this temperature. This amount of concentration soon turns to supersaturate when the sample is cooled. Some are diffused out of the crystal, when the cooling rate is slow enough for the vacancies to escape from the crystal. However, at temperatures far lower than $T_{m}$, outward diffusing cannot occur, as diffusivity is almost negligible.
On one hand, vacancies can be introduced by electron- or ion-beam bombardments at low temperatures. These created vacancies remain unless annealed by raising the temperature. The vacancy concentration $c_{d}$ behaves virtually like an independent variable against $T$ and $V$. In the semiconductor industry, indeed, the dopant concentrations are almost regarded as controlling parameters, similar to temperature.

The silicon crystals used in electronic devices are the closest-to-perfect crystals produced by humans. Dislocation-free crystals were achieved by a special technique \cite{Dash58}. Yet, intrinsic point defects remain, such as vacancies and interstitials. Engineers working on Si wafers came to recognize more seriously the presence of these intrinsic defects after dislocation-free crystals were achieved. Dislocations serve as sinks that trap these point defects: without dislocations, vacancies and interstitials are uniformly spread out over a wafer. This fact already indicates that the equilibrium concentration of defect hinges on the local structure. 

We can say further. In silicon crystal growth from a melt, the concentrations of vacancies and interstitials are known to have dependence on the pulling rate $v$ of crystal at $T_{m}$: vacancy-rich (interstitial-rich) crystals are obtained at a high (low) $v$ \cite{Voronkov82,Voronkov99}. The properties of the obtained crystal thus have the rate dependence. During cooling to room temperature, the concentrations of vacancies and interstitials have spatial distribution because of the finite time of diffusion \cite{Vanhellemont11,Voronkov12}. Thus, the manufactures of Si wafers guarantee the same wafer only within the tolerance of the defect concentrations. One cannot state that obtained properties of Si wafers are determined uniquely by $T$ and $V$ solely.

Importantly, point defects often develop further to form defect complexes and secondary defects. It is considered that A-type defect (swirl defect) is formed from interstitials \cite{Abe66,Foll75} and D-type is formed from vacancies \cite{Roksnoer81}. Also, formation of different types of microvoids are reported during the complicated processes of semiconductor devices \cite{Ryuta90,Itumi95}. These voids are considered as the precipitation of vacancies. Often, point defects are more stabilized by agglomeration. It is impossible to say the equilibrium configuration of silicon unique to $T$ and $V$. 

In metallurgy, it is daily experience to observe a variety of microstructures, whose morphology spans from lamellar, plate like, dendritic, acicular, to globular. There is no ``equilibrium" morphology to a specific $T$. The obtained morphology highly depends on the past history, mostly on the growth kinetics of nucleation \cite{Shewmon}. After forming, the morphology almost does not change anymore. 
It is hopeless task of finding the global equilibrium state among these microstructures, Despite this, it is quite legitimate to assign a unique value $U$ for each microstructure by specifying $\{ \bar{\bf R}_{j} \}$, as indicated by Eq.~(\ref{eq:UofSolids}). Each microstructure has its own FRE, by regarding it as an equilibrium state. As seen in the next subsection, these microstructures determine the mechanical properties of solids. Indeed, it is more fruitful to admit that all metastable states are equilibrium states within their relaxation times and to assign $U$ value to them. The same holds for glass too: all the existing glass samples are equilibrium states as long as their structures do not change.

\subsection{Mechanical properties and aging}
\label{sec:Mechanical}
It is known that the mechanical strength of metals is not sensitive to crystallographic structures but is strongly affected by defects. Majority of plastic deformations are mainly mediated by dislocations. Yield stress is determined by the balance between dislocations and the local obstacles to dislocation motion \cite{Cottrell67}. Interstitial atoms often play the role of these obstacles; a typical example is steel.
For dislocations, there is no idea of the ``equilibrium" concentration. The dislocation concentration is determined by the details of sample preparation, and is affected by working history (work hardening). From this observation, one can conclude that the mechanical properties of all solids have history dependence and cannot be uniquely determined by $T$ and $V$ solely \cite{Berdichevsky06}. 

\begin{figure}[htbp]
    \centering
    \includegraphics[width=75 mm, bb=0 0 650 420]{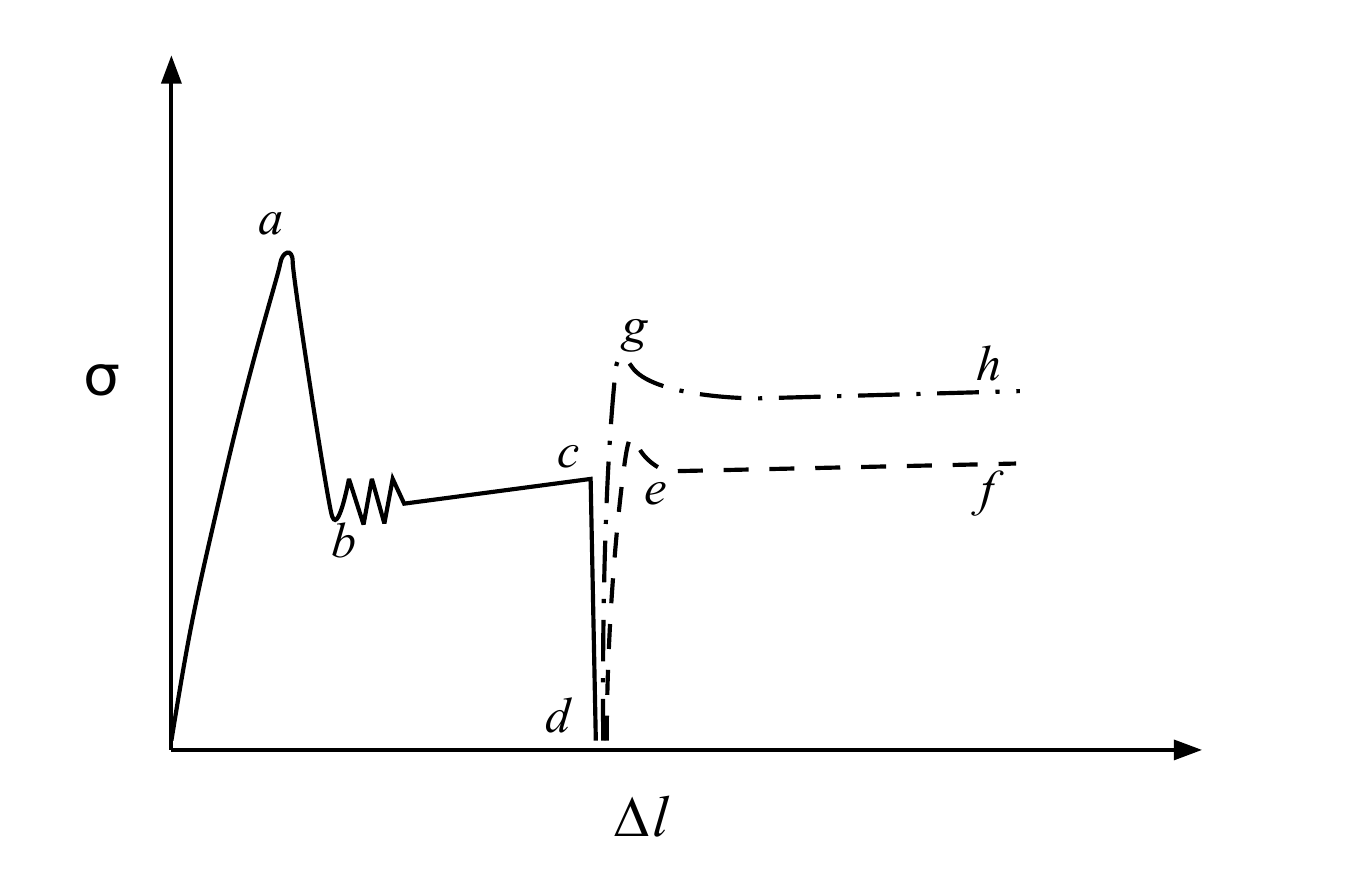} 
  \caption{Tensile test on a steel specimen. Stress $\sigma$ is plotted against the elongation $\Delta l$ of a test piece \cite{KodaNariyasu73}. States $a$ and $b$ correspond to the upper and lower yield points, respectively.
  } \label{fig:strain-aging}
\end{figure}

Aging is a special mode of history-dependent phenomena. Figure \ref{fig:strain-aging} illustrates a tensile test, in which the plastic deformation of a steel sample is measured. A test piece with the length $l$ is subjected to a load. For the elastic regime, where the load is small, the elongation $\Delta l$ vanishes when the load is removed. After the load is increased beyond the yield point ($a$ in Fig.~\ref{fig:strain-aging}), dislocations begin to move. At some point $c$, let us stop applying the load. $\Delta l$ never returns to zero, and the test specimen stays at an elongated state $d$. Here, plastic deformation occurs. If loading is resumed immediately after stopping, the elongation begins to trace the same path as that before stopping: $e \rightarrow f$ in Fig.~\ref{fig:strain-aging}. If loading is restarted after a waiting time $t_{w}$, the specimen undergoes a different path: $g \rightarrow h$ in Fig.~\ref{fig:strain-aging}. This effect is called {\it strain aging}. 
For low-carbon steel, it is known that interstitial C atoms accumulate near the place of dislocations during the waiting time $t_{w}$. The resulting C precipitation locks the movement of dislocations \cite{Cottrell67,KodaNariyasu73,Shewmon}. Because this process highly depends on the diffusion of the interstitial C atoms and their local distribution, the exponential-type time variation or the Arrhenius type variation, Eq.~(\ref{eq:rate}), cannot be expected.

Once we know the mechanism of strain aging in this manner, there is no surprise to find this effect in other crystals, no matter how small the effect is, because dislocations and interstitial atoms commonly exist in all crystals. The notion of dislocation does not apply to glasses. However, for metallic glasses, the role of dislocation is played by the so-called shear transformation zone \cite{Argon77,Spaepen77}. Similar mechanisms are reasonably expected for network glass through analogous media; some regions have higher locally strain than others.
Various aging effects have been reported for a wide range of glass materials, depending on their properties \cite{Mossa04,Fukao05,Vincent07,Amir12,Mamiya12,Samarakoon16,Peyrard20}. 
Although the author does not know the mechanisms for individual cases, it is certain that all mechanisms are common at involving the existence of many local minima in the structures and their relaxation times.

Regarding dislocation motions, a general consensus is obtained that the activation mechanism is in a broad sense thermally activation \cite{Schock80,Cailard-Martin03}. However, there is no unique energy barrier $E_{b}$ associated with a particular crystal. $E_{b}$ depends on the arrangement of local obstacles around a dislocation. Therefore, there is no simple formula for the yield stress. This contrasts to impurity diffusion, in which $E_{b}$ is almost uniquely determined when the impurity species is specified. The dynamic behavior of dislocations changes its modes in different $T$ ranges. Four modes, from the inertial motion to dragging motion, are identified depending on $T$ \cite{Cailard07}. It is unsuitable to adapt a simple transition rate, Eq.~(\ref{eq:rate}), with a single parameter $E_{b}$ to this problem. Similarly, for long-term changes such as degradation, it is common that the total change consists of several stages; The aqueous corrosion of glass is such an example; it consists of three stages: hydrolysis, precipitation, and secondary precipitation \cite{Frankel18,Ojovan20}.
For the glass transition, experimentalists found non-Arrhenius behavior for the temperature dependence of viscosity $\eta$ \cite{Angell95,Angell00,Lubchenko07}. Because the glass transition is the structural change occurring in a narrow range of temperature, it is reasonably expected that the energy barrier for viscous motions has a strong $T$ dependence \cite{Debenedetti01,Dyre04,Han20,Shirai21-GlassHysteresis,Xue22}. In this situation, it is demonstrated that the conventional analysis of Arrhenius plot gives large overestimations for the energy barrier \cite{Shirai21-ActEnergy}.

\subsection{Summary of history-dependent phenomena}
\label{sec:SumHysteresis}
Up to here, we have seen that hysteresis phenomena are very common properties that can be observed in any solid if studied at a sufficient resolution. By adapting the nonequilibrium reasoning of glass to other solids merely because of hysteresis, the destructive conclusion that all solids are in nonequilibrium states would be deduced. The history-dependent behaviors of solids are due to the existence of many configurations $\{ K \}$ and their relaxation times $\tau_{K}$
The difficulties regarding equilibrium can be resolved by specifying the states of solids using Eq.~(\ref{eq:StateOfSolid}). All states of solids are well described through thermodynamic methods in a certain time of period. 
Our heuristic verification of Eq.~(\ref{eq:StateOfSolid}) consists of the following steps.
\begin{enumerate}
\item{Any solid has a large number of metastable configurations $\{ K \}$, depending on the time period of observation.}

\item{Each configuration $K$ retains its structure, specified by $\{ \bar{\bf R}_{j}^{K} \}$, within the relaxation time $\tau_{K}$. In this period, the configuration $K$ can be regarded as equilibrium.
}

\item{Given a set of the equilibrium positions $\{ \bar{\bf R}_{j} \}$, the internal energy $U$ can be uniquely determined by Eq.~(\ref{eq:UofSolids}). This is consistent with Postulate \ref{PS:internal-energy}.
Moreover, this consistency is established only when the set $\{ \bar{\bf R}_{j} \}$ is considered the state variables for solids.}

\end{enumerate}

All the ingredients used in this verification are experimental facts and Postulate \ref{PS:internal-energy} only.
The important thing is that all equilibria are effective only in a certain time period. In the elemental course of thermodynamics, time is out of consideration. Although this treatment is appropriate for the elemental course, real studies inevitably encounter the time issue, as experiments are performed in certain time periods. In Section~\ref{sec:Equilibrium}, we appraise the definition of equilibrium from this point of view.

\subsection{Approaches from nonequilibrium thermodynamics}
\label{sec:OtherApproaches}
Before proceeding to the argument on thermodynamic foundation, a comment is given on the traditional approach to history dependent phenomena in nonequilibrium thermodynamics.
In the literature, history dependent phenomena are sometimes described by introducing effective temperature $T^{\ast}$: see \cite{Nieuwenhuizen98a,Bouchbinder09-2,Sciortino05,Mezard12} also \cite{Casas-Vazquez03} for review. These include the fictive temperature \cite{Tool31,Tool46,Davies53a}, which is used in the glass literature. Their attempts are regarded as replacing $\{ \bar{\bf R}_{j} \}$ in the full expression (Eq.~(\ref{eq:UofSolids})) by a single parameter $T^{\ast}$ as
\begin{equation}
Y = Y(T, V, T^{\ast}).
\label{eq:effectT}
\end{equation}
Despite their different historical backgrounds and nuances, the notions of internal variables and {\em order parameters} share the same intention to describe history dependent properties by extending variables \cite{deGroot-Mazur, Prigogine67, Davies53a}.
When a specific property $Y$ is focused on, this replacement may work. However, success cannot be guaranteed in general. Sciortino admits that a single parameter $T^{\ast}$ is insufficient to describe aging phenomena \cite{Sciortino05}. Then, the natural question is how many are required. Ritland pointed out that there are as many order parameters as the number of properties \cite{Ritland54-56}. In the present context, order parameters and internal variables for solids are equivalent to state variables \cite{Shirai23-OrderParams}. Equation (\ref{eq:effectT}) is only deemed an approximation of the exact statement (\ref{eq:StateOfSolid}).

\section{Equilibrium versus infinitesimally slow process}
\label{sec:Equilibrium}

\subsection{Failure of a previous Postulate}
\label{sec:FailureEq}
In this section, the fundamental statement (\ref{eq:StateOfSolid}) is derived from the thermodynamics laws.
The definitions of equilibrium and state variable are fundamental problems in thermodynamics. 
These two problems constitute a twin problem. State variables, such as temperature, are quantities that characterize equilibrium. However, the former is defined only when the latter is established. The definition is circular, similar to the chicken-and-egg problem \cite{Beauragard-note}.
Let us check how this issue is treated in an authoritative textbook of thermodynamics by Callen \cite{Callen}. In Sec.~1.5 of his book, this issue is discussed. He recognized this circular argument (\cite{Callen}, p.~15). Rather than derivation, he adopted an approach based on postulations. The correctness of postulation is assessed by the result that it leads to. He put the following postulate forth.

\vspace{3mm}
\noindent
{\bf Callen's equilibrium postulate}:
{\em There exist particular states (called equilibrium states) of simple systems that, macroscopically, are characterized completely by the internal energy $U$ and the volume $V$.} (\cite{Callen}, p.~13)
\vspace{2mm}

\noindent
The original statement of Callen includes additional variables of chemical compositions are included, but these are omitted here because chemical reactions are not discussed at this moment. Through an appropriate transformation of variables, this statement is read as the equilibrium state of a system is uniquely specified by $T$ and $V$. This is equivalent to the statement of Eq.~(\ref{eq:StateOfGas}). This postulate is correct for gases. Although its applicability to solids is uncertain, he generalized it to solids.
He recognized the history-dependent behaviors of solids, such as the hysteresis of steel. At this moment, he had two choices to proceed with: either that his postulate is wrong for solids or that steel is not an equilibrium material. He chose the latter conclusion. Whether Callen considered steel as a rare case is uncertain. However, we have already seen in Sec.~\ref{sec:Hysteresis} that hysteresis behavior is highly common across solids. Hence, if Callen's view is adopted, then all existing states of solids would turn to be in nonequilibrium states. This is a dilemma surrounding solids in thermodynamics.

A repeatedly claimed explanation to escape from this difficulty is to defer the establishment of equilibrium for future. It is often claimed that if we wait for a long time, equilibrium will be reached.
By observing that memory effects eventually fade away, Callen explained that these nonequilibrium states will finally reach equilibrium. This is sometimes true. Defects introduced by ion bombardment can be removed by annealing at high temperatures until such defects can be practically ignored. But complete removal is impossible. A finite defect concentration $c_{d}$ remains. This residual concentration is the equilibrium concentration $c_{d}(T_{a})$ at the annealing temperature $T_{a}$. This cannot be reduced further by thermal treatment only. The residual concentration varies depending on defect types. As defect study is developed, new defect types will be found. To discuss which defect type is the most stable equilibrium state may be meaningless. 
In addition, as discussed in Sec.~\ref{sec:DefectConcentration}, point defects interact each other and sometimes to form complexes. For example, human-made diamond (type Ib diamond) contains high concentrations of point defects of nitrogen impurities; these nitrogen impurities form a complex N$_{4}$ at long-time annealing at high temperatures instead of removing \cite{Jones01}. Dislocations and grain structures are hardly removed by annealing. When two configurations $K$ and $K'$ are close in the configurational space, the transition can occur. But when these are very far apart, as is the case of states $a$ and $d$ in Fig.~\ref{fig:meta-frozen}, the transition is almost inhibited. Callen's Postulate on equilibrium does not meet experimental facts.

Over a century, it has been said that a glass eventually becomes a crystal or the so-called ``ideal" glass (the state with zero residual entropy). Actually, this has never been observed, as stated in Introduction \cite{Berthier16}. Still one can stick with this view, by saying that our waiting time is not enough. No one can prove nor disprove this. The conventional view has thus been protected by this impossibility of disproof.
\cite{note-expectation}
However, this conventional view involves an expected future event. It is not advisable to use expectation for constructing physics laws. If we wait too much, unexpected stable states are sometimes discovered, as the example of $\delta$-orthorhombic boron indicates. Point defects often further develop to form a new type of defect, as the formation of microvoids from vacancies in silicon indicates. The domain structure of magnetism may be further developed after a long-time waiting after removing an external field. Waiting for an astronomical time may be even worse. Nuclear reactions will occur. Researchers are not quite sure whether protons have a finite lifetime. Even astrophysicists do not know the fate of our universe, whether the Big Crunch occurs or not.

\subsection{Structure is akin of constraint}
\label{Sec:PassiveResistance}
The most difficult problem about equilibrium is the distinction between no change and infinitesimally slow change, which causes many confusions. By observing aging effects in glass, it is often claimed that the current state of glass is changing very slowly. However, aging is a common phenomenon in solids, as already discussed in Sec.~\ref{sec:Hysteresis}. Aging is essentially atom relaxation. Scherer recommended using the term ``structural relaxation" to refer to aging \cite{Scherer84}.

The issue of infinitesimally slow process in thermodynamics is not new. More than a century ago, Gibbs studied this problem, when the criteria of equilibrium are discussed (\cite{Gibbs-TD}, p.~56; see also \cite{Commentary-Gibbs-Wilson}, p.~35; \cite{Commentary-Gibbs-Butler}, p.~74). It is unfortunate that his study was not well reflected in thermodynamics textbooks yet. By analyzing chemical reactions, he recognized two kinds of static states of materials. The first one is a static state achieved by the balance between two active tendencies with opposite directions. This corresponds to the usual chemical equilibrium, $A \leftrightarrow B$, where the forward and backward reactions are counterbalanced. The second one is a static state of no reaction: $A \not \rightarrow B$. For example, consider the reaction of ammonia production, 
\begin{equation}
{\rm N_{2} + 3 H_{2} \rightarrow 2 NH_{3}}. 
\label{eq:ammonia}
\end{equation}
Standard thermochemical data show that the free energy change of reaction (\ref{eq:ammonia}) at room temperature is $\Delta G = -16.2$ kJ/mol (\cite{Yazawa-collection11}, p.~92). The negative sign dictates that the reaction (\ref{eq:ammonia}) should occur in the forward direction. In reality, however, no reaction occurs when these two gases are mixed at room temperature. A substance inhibiting this chemical reaction (\ref{eq:ammonia}) is required. This substance was called ``passive resistance" by Gibbs. When he formulated his criterion for equilibrium, he adapted the same criterion to both cases by acknowledging passive resistance. On this understanding, Gibbs' criterion for equilibrium becomes equivalent to Postulate \ref{PS:second-law}.
Later, passive resistance was alternatively called inhibitor, anti-catalysis, etc, in various areas \cite{Hatsopoulos}. From the contemporary viewpoint, the name {\em constraint} may be the most preferable name. 

In the Gibbs' era, the nature of passive resistance was not known. No atomic theory existed. Today, we know that the substance of passive resistance is the energy barrier. Now, we do not need to distinguish the above two cases. No equilibrium is free from constraints. For the first case, the wall of a container is implicitly assumed in order to specify the volume $V$ of the gas. This is the constraint for the first case. Microscopically, energy barriers are built between the atoms in the metal wall, which impede the gas molecules from diffusing out. In reality, even the robust metal walls of gas cylinders cannot retain their initial pressure for long years. Because the energy barrier has a finite height, gas molecules diffuse at a very slow rate. As for ammonia production, this reaction occurs in the revolution of the sun system. (The author does not know whether the ammonia production was the consequence of long-time waiting or the consequence of other external impacts, such as explosion.)
Any kind of fossil fuel must be in a stable equilibrium state within its relaxation time: otherwise it would be exhausted before use. In this manner, we see that every equilibrium state is sustained by constraints, and no equilibrium is eternal.

In classical thermodynamics, the notion constraint can be easy understood by imagining a wall-like structure (\cite{Reif}, p.~87). The gas molecules in a container cannot escape from the container due to the presence of the wall. The wall is thus a constraint $\xi_{1}$. If a container has $m-1$ internal walls, the container is separated into $m$ compartments. The whole system is specified by the walls $\{ \xi_{j} \}_{j=1, \dots , m}$.
The walls are visible in this case. However, this role of constraint inside materials is not easy to imagine, as the energy barrier is invisible. Actually, from the microscopic viewpoint, inhibition of escape from the container (gas case) and inhibition of atom movement beyond a unit cell (crystal case) are the same. Both involve potential barriers that inhibit atom movements. In analogy to a container with many compartments, the boundaries of unit cells can be regarded naturally as as constraints $\xi_{j}$.

\begin{Definition}{\rm (Constraint)}
A constraint $\xi_{j}$ is an object to inhibit the change in $j$th-motion of the constituent particles of a system.
\label{def:constraint}
\end{Definition} 
A system is characterized by a set of constraints $\{ \xi_{j} \}$ (\cite{Gyftopoulos}, p.~12). The set of constraints $\{ \xi_{j} \}$ of a system determines its atom configuration $K$ for solids. The constraints of a system are thus the structure of the system.

Constraint $\xi_{j}$ accompanies the barrier height $E_{b,j}$, which is always finite. This determines the relaxation time $\tau_{j}$ (Eq.~(\ref{eq:rate})). Therefore, the lifetime, $\tau_{K}$, of configuration $K$ is determined by the shortest among various relaxation times $\{ \tau_{j} \}$.

\begin{Corollary}{\rm (Equilibrium period)}
The atom configuration $K$ of a material does not change in a time period $\tau_{K}$, which is determined by a set of constraints $\{ \xi_{j} \}$. This state is in equilibrium in this time period $\tau_{K}$.
\label{Co:period}
\end{Corollary}
For a given crystal, many microstructures can be realized by creating defects. Conventionally, defect states are not regarded as equilibrium states; instead, they are classified as metastable states. However, all defect states must be equilibrium states in their relaxation times $\tau_{K}$. The defect-free crystal is not thermodynamically stable. Depending on the energy barrier, $\tau_{K}$ can be any value from microscopic time to almost infinity. One millisecond is short in our timescale but is still sufficiently longer than the temperature relaxation time (of the order of 1 ps) of solids, as discussed in paragraph $f$ in Sec.~\ref{sec:Terminology}.  Given a well-defined $T$, these defect states are equilibrium states within their relaxation times. Corollary \ref{Co:period} may be valid even beyond the category of solids. For $\alpha$-radioactive nuclei, the lifetime spans an extraordinarily wide range of $10^{-6}$ s to $10^{10}$ years (\cite{BerkleyPhysics-4}, in paragraph 37 of Chap.~7). It is today's understanding that the lifetime is governed by the energy barrier for $\alpha$ particles to escape from nuclear forces: however the transition mechanism is the quantum tunneling. Though isotope U$^{238}$ is a stable nuclide, it will be decayed in $10^{10}$ years.

When special words aging and memory effects are mentioned, one is liable to consider that only special groups of materials possess these effects. However, as already discussed in Sec.~\ref{sec:Mechanical}, all these phenomena represent the competitive behaviors of equilibrium and relaxation among many metastable states, and they occur in all solids, irrespective of the magnitude of the effect. (Of course, for industrial applications, only selective materials with large effects qualify as memory materials.)
Characterizing a material as nonequilibrium merely because of observing aging effects is inappropriate in the thermodynamics context. By observing that the crystallographic structure did not change in the X-ray diffraction (XRD) measurement, one concludes that a silicon crystal is an equilibrium material. XRD is insensitive to defects, unless the amount of defects is too large. However, we observe a long-term inter-diffusion of dopant atoms at $pn$ junctions of semiconductor devices, which eventually limits the device performance. Identification of equilibrium state depends on the scope of the problem; that is, it depends on which variables among many $\bar{\bf R}_{j}$ are in focus.

\subsection{Consistent definition of equilibrium}
\label{Sec:Equilibrium-Def}

Corollary \ref{Co:period} already uses the term state variable, because the configuration is specified by $\{ \bar{\bf R}_{j} \}$. Hence, this cannot be the solution to the fundamental difficulty of the circular argument between equilibrium and state variables, although the statement itself is correct. Gibbs' postulate for equilibrium is also not free from this fundamental difficulty: his postulate makes sense only when the state variable of entropy is defined beforehand. This difficulty was solved by Gyftopoulos and Berreta \cite{Gyftopoulos, Shirai18-StateVariable}. The failure of all the previous attempts to define state variables lies at the self-contradiction that the notion equilibrium is explained by the word equilibrium. Gyftopoulos and Berreta started from the general situation, which is not in equilibrium and does not involve the idea of state variables. Then, they figured out that equilibrium is a special case of nonequilibrium states.
\begin{Definition}{\rm (Equilibrium)}
it is impossible to extract work from a system in equilibrium to the outside without leaving any effects on the environment. {\rm (\cite{Gyftopoulos}, p.~58)}  
\label{def:equilibrium}
\end{Definition} 
This definition does not mention state variable. Furthermore, this definition is favorable in that no information about the internal structure is required. It only uses external effects that can be observed.
When gas turbulence occurs in a container, work can be extracted from it. When a chemical reaction occurs in a system, work can be extracted by inserting a heat engine to convert the generated heat into work.
If work were extracted from the equilibrium system without affecting the environment, the only possible change in the energy is a decrease in the internal energy $U$ of the system. This means that work was obtained by cooling solely. This contradicts the well-known Clausius statement of the second law. In this manner, we see that there is no exception for Definition \ref{def:equilibrium}.
It is our daily observation that work cannot be extracted from glasses. Hence, the only conclusion from this observation is that glass material must be in equilibrium, as long as their atom configurations do not change.

\subsection{Invariance property of state variables to thermal fluctuations}
\label{Sec:StateVariable-Def} 
Now that we have defined equilibrium without using state variables, it is logical to define state variables by using equilibrium. Equilibrium means no change. This raises the question of which variable does not change. Microscopically, physical properties have time dependence due to thermal fluctuations. The volume of a container fluctuates microscopically, $V(t)$. The wall of the container comprises atoms, whose positions fluctuate. However, when the system is in equilibrium, the time average of volume can be defined. We know that the time-averaged value $\bar{V}$ is a state variable. Even when equilibrium is established, the instantaneous value of the density of a gas in the container has microscopic thermal fluctuation: $\rho(t)=\bar{\rho} + \delta \rho(t) $. However, by taking time average, the mean density $\overline{\rho(t)}$ does not change over time: in more precisely, we say that the time average does not depend on the period on which the time average is obtained. 

We can generalize this procedure to define state variables.
Let us consider a dynamical variable $X(t)$ that is subject to a certain constraint $\xi_{j}$. 
The constraint $\xi_{j}$ on $j$th motion means the imposition of a range of permissible changes in $X(t)$. 
\begin{Definition}{\rm (State variable)}
If the time average $\overline{X(t)}$ has a definite value when the system is in equilibrium, then the state variable of $j$th type, $X_{j}$, is defined by 
\begin{equation}
X_{j} = \frac{1}{t_{0}} \int_{\xi_{j}} X(t) dt,
\label{eq:jth-SV}
\end{equation}
where $t_{0}$ is the period of the time average. \label{def:SV}
\end{Definition}  
By the definite value, it is meant that the time-averaged value is independent of the time period $t_{0}$, in which the time average is obtained. For example, for a gas in a box with volume $V$, the constraint is the wall of the box. The position, $x(t)$, of a gas molecule is constrained within the wall. In equilibrium, the molecule visits everywhere in the box at equal probabilities. The time average of the molecule's position, $\overline{x(t)}$, is indeterminate. Thus, $\overline{x}$ is not a state variable, as is leaned by the elemental course of thermodynamics. Now consider the spatial distribution function, $\delta(x', x(t))$, of a molecule at $x'$. The integration of this function over the inside of the box
\begin{equation}
X(t) = \int_{V} \delta(x', x(t)) dx',
\label{eq:int-distribution}
\end{equation}
is interpreted as the density-weighted volume. When the molecule distribution is uniform in space, which is obtained in equilibrium, the time-averaged value $\overline{X(t)}$ gives the volume $V$. The time-averaged value $\overline{X(t)}$ becomes a state variable $V$. The nonvanishing property after time averaging is the most important property of state variables, and we may call this {\em the invariance to time averaging} 
or {\it the invariance to thermal fluctuations}. 

A thermodynamic system consists of many constraints $\{ \xi_{j} \}$. The wall of a container determines the shape of the gas in the container. At a fixed temperature, only one independent state variable $V$ exists. When a mobile wall is inserted into this container, the structure of the whole system is specified by two volumes $V_{1}$ and $V_{2}$, separated by the internal wall. The state variables are these two volumes $V_{1}$ and $V_{2}$. This partitioning can be continued to any number of compartments. One new state variable is created each time a constraint is inserted. This one-to-one correspondence between constraints and state variables was clarified by Reiss \cite{Reiss}. The structure of a system is fully specified by the set of constraints $\{ \xi_{j} \}$, and thus the equilibrium properties of the system are completely determined using the full set of corresponding state variables $\{ X_{j} \}$.

The heuristic finding of this one-to-one correspondence between constraints and state variables is proven rigorously from the second law of thermodynamics expressed in Postulate \ref{PS:second-law}.
This guarantees that the state variable $X_{j}$ is uniquely determined for a constraint $\xi_{j}$ ({\em uniqueness}). As stated in Definition \ref{def:SV}, the uniqueness is the definiteness that the state variables must satisfy. For example, the time-averaged positions of atoms in a gas do not have unique values; hence, they cannot be state variables.
Moreover, Postulate \ref{PS:second-law} claims that equilibrium is uniquely specified by a set of constraints $\{ \xi_{j} \}$ ({\em completeness}).

By defining state variables rigorously, we find that the equilibrium positions of atoms in solids are state variables. Atom $j$ in a solid is so constrained that it can move only within a small region in a unit cell $\xi_{j}$. It has a unique value $\bar{\bf R}_{j}$ in equilibrium. Therefore, we conclude
\begin{Corollary}{\rm (State variables of solids)}
The state variables of a solid are the time-averaged positions, $\bar{\bf R}_{j}$, of all atoms comprising the solid.
\label{Co:atom-position}
\end{Corollary} 
The full set $\{ \bar{\bf R}_{j} \}$ of a solid uniquely determines the structure of the solid. This means that completeness is satisfied by this set. Here, we establish that the state function $U$ of solids is determined by Eq.~(\ref{eq:UofSolids}).

The expression (\ref{eq:UofSolids}) indicates that a large number $3N_{\rm at}$ of state variables are required for solids. This is the most contrasting thermodynamic property of solids relative gases, in which only volume $V$ is a state variable. Probably, this is the most difficult point from the traditional understanding of thermodynamics. It is the tenet of traditional thermodynamics that thermodynamic properties of macroscopic systems are described by a few state variables \cite{Callen}.
However, how large is macroscopic and how small is microscopic are meaningful only in the relative sense. We often distinguish them on a human scale. This manner is subjective, and unsuitable for describing physics. Atom positions are microscopic on a human scale. However, we can observe it, such as through XRD. The diffraction pattern exposed on a film is a macroscopic object, and the observed angles are a macroscopic quantity. If the mechanism of XRD is not considered, we might recognize that the crystal structure is characterized by a macroscopic observation.
Also, it is widely believed that the number of state variables must be small, which cannot accept the Avogadro's number of variables. However, again, it is the relative matter to say how big a large number is and how small a small number is. Infinity is a large number. However, there are many infinities. For example, the number of integers is infinity, but it is quite smaller than the number of real numbers. By definition, an average $\bar{\bf R}$ presumes the existence of a large number of microscopic states ${\bf R}(t)$. The essential property for state variables is the invariance to time averaging. The instantaneous positions ${\bf R}(t)$ at a moment are eliminated by the operation of time averaging. State variables survive after this operation; in other words, they are invariant to time averaging. This invariance agrees with Callen's view (\cite{Callen}, p.~8). 
It is worthwhile to point out that the notions of internal variables and order parameters share the same idea of the invariance to time averaging for state variables. Hence, these quantities are regarded as essentially the same as state variables \cite{Shirai23-OrderParams}.

Glass transition exhibits dependence on the cooling rate $\gamma$. For this reason, it is often claimed that deducing thermodynamic properties of material from the observed $C_{p}-T$ curve has no sense \cite{Hodge94}. This is false \cite{Shirai20-GlassState,Shirai21-GlassHysteresis}. During the transition, the shape of $C_{p}-T$ curve can vary depending on $\gamma$. However, the final state of glass has the definite values for the time-averaged positions $\{ \bar{\bf R}_{j} \}$, which specify uniquely the state of the glass. If $U$ were not uniquely determined by $\{ \bar{\bf R}_{j} \}$, $U$ could have multivalues and the energy conservation would be violated by Postulate \ref{PS:internal-energy}.
Glass transition is identified by the region at which a jump, $\Delta C_{p}=C_{p}^{l}- C_{p}^{g}$, occurs: $C_{p}^{l}$ and $C_{p}^{g}$ are the specific heat of the liquid and glass states, respectively. Because $\Delta C_{p}$ is given only by the terminal states of the transition process, the jump $\Delta C_{p}$ must be a thermodynamic property of glass; that is, it should not depend on the process. The obtained structure $\{ \bar{\bf R}_{j} \}$ has certainly dependence on the cooling process. However, this merely indicates that various structures are obtained by changing the preparation conditions. Once the structure was fixed by $\{ \bar{\bf R}_{j} \}$, the properties of obtained glass are determined by $\{ \bar{\bf R}_{j} \}$ solely. If two samples prepared by different processes happened to have the same structure $\{ \bar{\bf R}_{j} \}$, they must have the same properties with the same state variables $\{ \bar{\bf R}_{j} \}$. 

Two comments can be added from recent studies on the glass transition. First, the glass transition temperature $T_{g}$ in experiment has well-characterized lower bound. Although $T_{g}$ certainly varies depending on the cooling rate $\gamma$, the range that $T_{g}$ can vary is narrow, in most cases only in a few K \cite{Bruning92}. After compiling reported values of $T_{g}$ for silica, Mazurin concluded that $T_{g}$ is converged if appropriate conditions for preparation are imposed \cite{Mazurin07}. This qualifies that $T_{g}$ is a property of a material.
Second, recent studies on $\Delta C_{p}$ by DFT molecular-dynamics simulations demonstrated that $\Delta C_{p}$ is determined by the structural change in $\{ \bar{\bf R}_{j} \}$ \cite{Shirai22-SH,Shirai22-Silica}. This is in contrast to the traditional view that the glass transition is a kinetic transition. The kinetic reasoning does not have quantitative explanation for the occurrence of specific-heat jump. Th kinetic energy does not exhibit a jump at $T_{g}$, and only the potential energy does it. These studies also show that the deviation of the Prigogine--Defay ratio from unity is a consequence that $\{ \bar{\bf R}_{j} \}$ are state variables for glasses.

\subsection{Reproducibility}
\label{Sec:reproducibility}
Lastly, reproducibility of glass is discussed. History dependence brings troubles to material preparation. If the properties of the target material are not well reproduced, commercial production will be difficult. Fortunately, in many cases, experimentalists obtain a good reproducibility even for those materials with history dependence. 
In today's commercial productions of memory materials, manufacturers are successful to produce these materials with sufficient accuracy to meet high-quality industrial demands.
\cite{note-restorability}
By applying the same time sequence of input $X$, a certain relationship $Y=Y(X)$ between the input condition $X$ and the output result $Y$ is repeatedly observed in a certain range of $X$, even though this relationship is not reproduced when the time sequence is altered.
Phase change materials exhibit multi-valued resistance, which can be controlled by different sequences of pulses of the input voltage \cite{Kuzum11,Tuma16}. Their good reproducibility for repeated inputs is remarkable.
Reproducibility is obtained in glasses too. By considering that no glass has the same microstructure in its details, it is surprising that the glass products having the same properties with high accuracy are served to high-technology industries. 

Behind the reproducibility of these commercial products, there may be tremendous efforts in industrial researches. The author does not know these secrets of technology. Here, he only mentions the thermodynamics principles underlying the reproducibility. There are two factors to contribute this. First, good reproducibility itself infers that the final state $A_{f}(X)$ for each input $X$ is an equilibrium state after sufficiently long-time waiting. If the final state were not stable against thermal fluctuation, reproduction of the same state would not be possible. 

Second, there is degeneracy in thermodynamic states $\{A_{j} \}$. We now know that any state of solids can be uniquely specified by the time-averaged positions $\{ \bar{\bf R}_{j} \}$. But the converse is not true. The same properties do not imply the same structure $\{ \bar{\bf R}_{j} \}$. There are many configurations $\{ \bar{\bf R}_{j}^{K} \}$ that produce the same properties. Experimentalists identify the state of a material by observing a set of a limited number of properties $\{ P_{j} \}$ ($j=1, \dots r$), where $r$ is the number of properties that the experimentalists used for identification of state. Use of only one property ($r=1$) is not rare. There are a large number of energy degeneracy for solids. For a single point defect, there are $N_{\rm s}$ sites available for the point defect. They cannot be distinguished by measuring $U$ solely. It creates the configuration entropy $S_{\rm conf} = k_{\rm B} \ln N_{\rm s}$. Degeneracy depends on the resolution of measurement on energy. Within this resolution, all configurations $\{ \bar{\bf R}_{j}^{K} \}$ of a glass are identified as having the same energy. If other properties $P_{j}$ which are independent of $U$ are used for distinguishing them, one can further classify the glass properties. In this manner, identification of individual glasses depends on the used set of properties $\{ P_{j} \}$ and their resolutions \cite{Shirai18-StateVariable}. This is the meaning what ``the same" glass indicates, the question which was posed in paragraph $d$ in Sec.~\ref{sec:Terminology}.

\section{Conclusion}
The glass state has long been believed as a nonequilibrium state, because of its history dependence. 
In this paper, this characterization has been assessed from the fundamentals of thermodynamics. This is needed because the previous arguments did not distinguish thermodynamic states from processes, which causes many confusions. The problem with previous arguments is the hypothesis that $T$ and $V$ are the only independent state variables. Although this is true for gases, its applicability to solids is never proven. In fact, experiments show that solids have an enormous numbers of microstructures, such as defects, grains, and metallurgical structures. These are metastable states: their equilibrium states last only within a finite time (the relaxation time). Traditionally, only one stable equilibrium state, which is unique to $T$ and $V$, is deemed to exist for a given solid. However, eternal equilibrium is noexistent. The perfect crystal is not even a local-minimum state in the free energy and hence is thermodynamically unstable. All metastable states must be in equilibrium within their relaxation times. 
The relaxation time is determined by the details of its microstructure, that is, not only the crystallographic structure but defects, morphologic, metallurgical structures. All these microstructures are specified by the time-averaged atom position, $\{ \bar{\bf R}_{j} \}$. Only by regarding $\{ \bar{\bf R}_{j} \}$ as state variables for solids, the property of state function of energy (Postulate \ref{PS:internal-energy}) can be recovered: otherwise, the energy conservation will be broken.
Any existing glass state is an equilibrium state as long as the structure does not change. The fact is that the relaxation time of silica glass may be the longest among various materials.

Numerous experimental facts indicate that history dependent phenomena are common across all solids, as plastic deformations indicate. The material properties of any solid can be affected by its preparation conditions. This is why studies on crystal growth exist for various solid. 
History dependence emerges from the fact that a hysteresis process consists of many intermediate states and that each state is indeed an equilibrium state. The intermediate structure is maintained in a finite time of period. This finite time delays the material response to the change in the external field, which brings about the history dependence.

\section*{Acknowledgment}
The author thanks Enago (www.enago.jp) for the English language review.


\begin{thebibliography}{140}
\bibitem[{not({\natexlab{b}})}]{note-svs}
\bibinfo{note}{In the following, any mention of $T$ and $V$ being the only
  state variables means that any state of a single-component system can be
  specified using a set of two independent state variables. Other state
  variables exist, such as pressure $P$ and entropy $S$. The essential matter
  is that gases have only two independent state variables among many. In this
  paper, a representative set of $T$ and $V$ is chosen for the two-dimensional
  state space. For anisotropic crystals, the trivial extension of state
  variables involves the use of a strain tensor in place of $V$. Magnetization
  and polarization state variables are used to describe magnetic and dielectric
  materials, respectively. These additional state variables are minor
  extensions of state variables to solids. This study focuses on a more
  fundamental extension.}
\bibitem[{not({\natexlab{c}})}]{note-hysteresis-nonEq}
\bibinfo{note}{There are a few approaches to treat hysteresis from
  nonequilibrium thermodynamics. Extended irreversible thermodynamics takes
  fluxes into account as fundamental variables \cite{Jou10}. Internal variable
  theory introduce additional freedoms of internal variables into theory
  \cite{Muschik93,Kestin70}. The most radical theory is rational thermodynamics
  \cite{Truesdell84,Hutter77}, viewing thermodynamic phenomena by a part of
  dynamics. All these theories study hysteresis by regarding it as the response
  to the previous influence. This view is correct, when the evolution of
  properties is studied. However, the main concern in this study is different:
  namely, how the present state is described within the framework of
  thermodynamics.}
\bibitem[{not({\natexlab{d}})}]{note-timeaverage}
\bibinfo{note}{The author deliberately uses the word of ``time-averaged
  position", instead of the simpler word ``mean position". For the glass
  transition, the distinction between a time average and an ensemble average is
  important.}
\bibitem[{not({\natexlab{f}})}]{note-restorability}
\bibinfo{note}{In material science, reproducibility is often used as a synonym
  of reversible. Bridgman used the word restorability by observing the same
  hysteresis curve in plastic deformations \cite{Bridgman50}. In the
  thermodynamics context, restorability is not the same as reversibility,
  because restoration in a hysteresis loop leaves its consequence on the
  environment. See Definition \ref{def:hysteresis}.}
\end{thebibliography}

\end{document}